\documentclass[twocolumn, times]{aastex631}
\usepackage{longtable}

\newcommand{\be}{\begin{displaymath}}
\newcommand{\ee}{\end{displaymath}}
\newcommand{\bea}{\begin{eqnarray*}}
\newcommand{\eea}{\end{eqnarray*}}

\shorttitle{Orbital Decay Revisited}
\shortauthors{Winn \& Stefánsson}

\graphicspath{{./}{}}
\usepackage{amsmath}
\usepackage{comment}

\usepackage{hyperref}
\hypersetup{
    colorlinks=true,
    linkcolor=blue,
    filecolor=magenta,      
    urlcolor=cyan,
    pdftitle={Overleaf Example},
    pdfpagemode=FullScreen,
    }

\begin{document}

\def\ltsima{$\; \buildrel < \over \sim \;$}
\def\lsim{\lower.5ex\hbox{\ltsima}}
\def\gtsima{$\; \buildrel > \over \sim \;$}
\def\gsim{\lower.5ex\hbox{\gtsima}}

\title{Orbital decay candidates reconsidered:\\
WASP-4\,b is not decaying and
Kepler-1658\,b is not a planet
}
\correspondingauthor{Joshua N.\ Winn}
\email{jnwinn@princeton.edu}

\author[0000-0002-4265-047X]{Joshua N.\ Winn}
\affiliation{Department of Astrophysical Sciences, Princeton University, 4 Ivy Lane, Princeton, NJ 08544, USA}

\author[0000-0001-7409-5688]{Guðmundur Stefánsson}
\affiliation{Anton Pannekoek Institute for Astronomy, 904 Science Park, University of Amsterdam, Amsterdam, 1098 XH, The Netherlands}

\begin{abstract}

The fate of hot Jupiters is thought
to be engulfment by their host stars,
the outcome of tidal orbital decay.
Transit timing has revealed a few systems
with apparently shrinking orbital periods,
but such signals can be mimicked by light
travel-time effects (LTTE) of a distant companion.
By combining transit timings with precise radial-velocity data,
including new data,
we reassessed three reported cases of orbital decay:
WASP-4, WASP-12, and Kepler-1658.
For WASP-4, the period change
is best explained by LTTE
due to a $\approx$$7~M_{\rm Jup}$ companion at $\approx$8~AU,
with no need to invoke orbital decay.
For WASP-12, in contrast, the data firmly exclude 
LTTE and confirm genuine orbital decay.
For Kepler-1658,
spectroscopic and photometric anomalies
reveal the ``planet'' to be an
eclipsing K/M binary bound to the F-type primary,
with LTTE explaining the observed period change.
Thus, among the known hot Jupiters, only WASP-12\,b
currently shows compelling evidence for orbital decay.

\end{abstract}


\section{Introduction}

Hot Jupiters have famously small orbits, some of which are only
3-4 times larger than their host stars.
These hottest of the hot Jupiters
are expected to be relatively short-lived on cosmological timescales,
slowly spiraling into their host stars as tidal interactions
sap their orbital energy and angular momentum.
In close stellar binaries, tidal interactions
usually lead to spin-orbit synchronization \citep{Hut1980},
but a hot Jupiter's angular momentum
is insufficient to spin a Sun-like star into synchrony
\citep{Rasio+1996, Levrard+2009},
especially since magnetized winds
steadily slow the star's rotation
\citep{BarkerOgilvie2009}.
However, the timescale of tidal orbital decay remains
uncertain because it depends on
poorly understood mechanisms
of tidal dissipation within stars, such as inertial-wave damping
and turbulent convection \citep[see, e.g.,][]{Ogilvie2014}.

Several population-level clues support
the idea that tidal decay does occur.
Sun-like stars with hot Jupiters rotate
faster, on average, than comparable stars without such planets
\citep{Brown2014,Maxted+2015,Penev+2018,TejadaArevalo+2021},
consistent with the tidal transfer of angular momentum.
Kinematic age studies show
that hot-Jupiter hosts are systematically younger
than matched field stars
\citep{HamerSchlaufman2019,MiyazakiMasuda2023}, implying
the planets are destroyed while
their stars are still on the main sequence.
More dramatically, the infrared transient
reported by \cite{De+2023} has been
interpreted as the aftermath of a giant
planet's engulfment.

Direct evidence for orbital decay has come
from transit timing.
The best studied case is WASP-12\,b,
whose transit period
is shrinking by about 30 ms/yr
\citep{Maciejewski+2016, Patra+2017, Yee+2020}.
However, secular changes in the transit period
can result from light travel-time effects (LTTE)
of a distant massive companion.
Also known as the ``R{\o}mer delay,''
the effect is familiar in
pulsar timing and triple-star systems \citep{Edwards+2006,Borkovits+2015}.
When a star and its transiting planet are being pulled
toward us, the observed transit period shrinks.
Because radial-velocity (RV) measurements directly track the
star's line-of-sight motion, they provide a useful test of
whether LTTE rather than orbital
decay is responsible for a shrinking transit period.

\cite{Yee+2020} used RV data to rule out LTTE
as the cause of WASP-12\,b's shrinking period.
Two other systems, WASP-4\,b and Kepler-1658\,b,
also have apparently shrinking periods.
Although light travel-time effects have been considered in both cases,
and some authors have argued that they have been
ruled out, no joint analysis of transit times and RVs has
yet been presented.

In this paper, we report new transit-timing and RV observations
of WASP-4\,b and Kepler-1658\,b, together
with joint analyses designed to test the LTTE hypothesis.
For illustrative purposes,
we also revisit WASP-12\,b,
for which strong arguments have already been
lodged against the LTTE hypothesis.
The paper is organized as follows.
Section~\ref{sec:math}
presents the relevant equations for the timing\,+\,RV analysis.
Section~\ref{sec:wasp-4} describes the case of WASP-4\,b,
and Section~\ref{sec:wasp-12} provides the contrasting
example of WASP-12\,b.
Section~\ref{sec:kepler-1658} describes our
investigation of Kepler-1658\,b and features
a plot twist:
we found evidence that
the transits are not caused by a hot Jupiter,
but rather by a low-mass eclipsing binary orbiting
a brighter and more massive star.
Brief reflections on the results are
given in Section~\ref{sec:discussion}.

\section{Mathematical preliminaries}
\label{sec:math}

The line-of-sight
coordinate displacement of a star from the center of
mass due to a single companion is
\begin{equation}
\label{eq:z}
z = \frac{a (1-e^2) }{1 + e\cos\theta } ~\sin I ~\sin(\theta + \omega),
\end{equation}
where $a$ is the semi-major axis,
$e$ is the eccentricity,
$I$ is the inclination,
$\omega$ is the argument of periapse,
and $\theta$ is the true anomaly.
The time dependence of $\theta$
is governed by
\begin{align}
\tan \frac{\theta}{2} &=  
\sqrt{ \frac{1+e}{1-e} }
\tan \frac{E}{2}~~{\rm and}\\
E - &e\sin E = n\left( t - t_{\rm p} \right),
\end{align}
where $E$ is the eccentric anomaly,
$t_{\rm p}$ is the time
of periapse passage,
and $n$ is the mean motion
($2\pi$ divided by the
orbital period $P$).
By taking the time derivative
of Equation~\ref{eq:z}, one obtains the
radial-velocity equation,
\begin{equation}
\dot z = \frac{2\pi a}{P \sqrt{1-e^2}} \, \sin I \left[\cos(\theta+\omega)+e\cos\omega\right].
\end{equation}
A few changes in parameterization
are useful.
First, we define
the radial-velocity semi-amplitude,
\begin{equation}
K \equiv \frac{2\pi a}{P \sqrt{1-e^2}} \, \sin I,
\end{equation}
because it can be measured directly
from radial-velocity data.
Second, instead of using $t_{\rm p}$ to pinpoint the
star's location at a particular time,
we use the true anomaly at inferior conjunction,
\begin{equation}
\theta_{\rm c} \equiv \frac{\pi}{2} - \omega,
\end{equation}
because it allows a more graceful
approach to the $e\rightarrow 0$ limit in which
$t_{\rm p}$ is ill-defined.
With these changes, $z$ and $\dot{z}$ can be written
\begin{eqnarray}
\label{eqn:z_exact}
z &=& \frac{K}{n}\left( 1 - e^2 \right)^{\!3/2}\,
\frac{\cos(\theta - \theta_{\rm c})}{1 + e\cos\theta},\\
\label{eqn:zdot_exact}
\dot{z} &=& K \left[e\sin\theta_{\rm c} - \sin(\theta-\theta_{\rm c})\right].
\end{eqnarray}

Assuming the star has a hot Jupiter
and a wider-orbiting companion
with negligible mutual interactions,
the observed radial velocity is
\begin{equation}
\label{eq:radial_velocity}
v = v_0 + \gamma + \dot{z}_1(t) + \dot{z}_2(t),
\end{equation}
where $v_0$ is the radial velocity 
of the planetary system's center of mass
relative to the Solar System,
$\gamma$ is a spectrograph-dependent additive
offset\footnote{The $\gamma$ parameter
is necessary because
although spectrographs can measure radial-velocity variations
with m/s precision,
the accuracy of the
absolute radial velocity scale is usually no better
than 100~m/s, leading to systematic offsets
between different spectrographs.},
and $\dot{z}_1$ and $\dot{z}_2$
are the contributions from the hot Jupiter and wider-orbiting companion, respectively.

In the same scenario, the
transit times are
\begin{equation}
\label{eq:transit_times}
t_k = t_{\rm c,1} + kP_1 + \frac{z_2(t_{\rm c,1} + kP_1)-z_2(t_{\rm c,1})}{c},
\end{equation}
where $t_{\rm c,1}$ is the time of conjunction of an
arbitrarily chosen orbit, $k$ is an integer,
and the last term is the LTTE correction.

It will be useful
to simplify these equations for the case when
the period of the wide-orbiting companion is
longer than the time span of the data. 
We choose $t=0$ to be
within the time span of the data,
and perform a Taylor expansion in the small parameter
$2\pi t/P_2 = n_2 t$:
\begin{equation}
\label{eqn:zdot_Taylor_expansion}
\dot{z}_2 = \sum_{j=0}^\infty \beta_j\,t^j =
\sum_{j=0}^\infty \frac{\beta_j}{n_2^j} \,(n_2 t)^j
\end{equation}
For the special case $e_2=0$, the
first four coefficients are
\begin{align}
\label{eq:beta_values_circular_orbit}
\beta_0=K_2\sin \phi_0,&~~\beta_1=-K_2 n_2\cos \phi_0 \\
\beta_2=-\tfrac{1}{2} K_2 n_2^2\sin \phi_0,&~~{\rm and}~~
\beta_3= \tfrac{1}{6} K_2 n_2^3 \cos \phi_0,
\label{eq:beta_values_circular_orbit_2}
\end{align}
where $\phi_0 \equiv n_2 t_{\rm c,2}$.
The corresponding approximation
for the transit times is obtained using
Equation~\ref{eq:transit_times} and the time
integral of Equation~\ref{eqn:zdot_Taylor_expansion}:
\begin{equation}
\label{eq:transit_time_quadratic_approximation}
t_k \approx t_0 + k\left(\!P_1 + \frac{\beta_0}{c} \!\right) +
\frac{1}{c}\sum_{j=1}^\infty \beta_j
\frac{(t_0 + kP_1)^{j+1} - t_0^{j+1}}{j+1},
\end{equation}
and the $\beta_0$ term is usually negligible.

\section{WASP-4}
\label{sec:wasp-4}

WASP-4\,b was the first planet discovered by
the SuperWASP-South
survey \citep{Wilson+2008}.
The host star is a G dwarf (5500\,K,
0.9\,$M_\odot$, 1.2\,$R_\odot$)
and the planet is a hot Jupiter
with an orbital period of 1.34~days,
a mass of 1.2$\,M_{\rm Jup}$,
and a radius of 1.4\,$R_{\rm Jup}$.

Whether or not orbital decay has been
detected for WASP-4\,b has been the subject
of debate.
Early timing studies
showed the transit
period to be shrinking
at a rate of order
$\delta P/P = 10~{\rm ms/yr}$
\citep{Bouma+2019, Southworth+2019},
although the robustness of this
finding was questioned \citep{Baluev+2019}.
Additional RV data suggested
the star is accelerating at a rate compatible with
the LTTE hypothesis \citep{Bouma+2020},
but concerns were raised
about systematic errors
\citep{Baluev+2020}.
With additional data,
\citet{Turner+2022}
confirmed a shrinking period
and argued that the LTTE hypothesis
was firmly ruled out,
but an error in their analysis was
later identified,
showing that LTTE was still viable
\citep{HarreSmith2023}.
Most recently, \cite{Basturk2025}
reported further transit times
and concluded that
orbital decay is the best explanation.

This sequence of claims and counter-claims
illustrates the difficulty of detecting
slight period changes and disentangling
orbital decay from LTTE.
We suggest that an underlying
reason for the confusion
is that none of the prior studies
performed a joint fit
of the radial-velocity
and transit-timing data.

\subsection{Radial Velocities}

Radial velocities spanning
13 years are available from
three spectrographs:
CORALIE\footnote{CORALIE is mounted
on the Euler 1.2\,m
telescope at La Silla Observatory, Chile.} \citep{Wilson+2008,Baluev+2019}, HARPS\footnote{High Accuracy Radial velocity Planet Searcher (HARPS)
is mounted on the La Silla 3.6\,m telescope.} \citep{Triaud+2010,Baluev+2019}, and HIRES\footnote{The High Resolution Echelle Spectrometer (HIRES)
was used on the Keck~I\,10\,m telescope on Mauna Kea, Hawaii.} \citep{Bouma+2020}.
We began with the compilation provided by \citet{Turner+2022}
and applied the following filters to construct
a clean dataset:

\begin{itemize}

\item We excluded all velocities measured during transits,
to avoid modeling the Rossiter-McLaughlin effect.

\item We omitted a single CORALIE velocity (BJD 2456149.87585)
with an unusually large uncertainty (59 m/s, compared
to the typical 15--25 m/s).

\item We added a new HIRES velocity from December 26, 2020 (BJD~2459189.71567), extending the HIRES time span
by 1.3~years. The full HIRES dataset is given
in the Appendix (Table~\ref{tbl:wasp4-rv}).

\item We ``thinned'' the data
by retaining only one point every five days.
This was done to avoid the
complexity of modeling time-correlated
noise.\footnote{\citet{Husnoo+2012}
fitted RV data with a model in which the correlation strength is a Gaussian function of the time interval between observations, and for WASP-4 the Gaussian width was 1.5 days. Thus, points separated 5 days are expected to have minimal correlations.}

\end{itemize}

The final dataset comprised
37 RVs: 23 from CORALIE,
5 from HARPS, and 9 from HIRES.
The data are displayed in the top left panel of Figure~\ref{fig:wasp4_data}.
The peak-to-peak
variations of about 400~m/s
are caused by the hot Jupiter.

\subsection{Transit Times}

WASP-4 was observed in
Sectors 2, 28, 29, 69, and 96
of NASA's Transiting Exoplanet Survey
Satellite (TESS) mission \citep{Ricker+2015},
and all the data are available with
120~s time sampling.
The Sector 96 data have not previously
been analyzed in the literature.
We used light curves
processed by the TESS
Science Processing and Operations
Center (SPOC) 
to measure transit times with the procedures
of \cite{IvshinaWinn2022},
obtaining 78 transit times 
with typical uncertainties
of 25~s. The new transit times
from Sector 96 are given in the Appendix (Table~\ref{tbl:wasp4-ttv}).

Many other transit times
have been reported for WASP-4\,b.
A challenge of long-term
timing analyses
is deciding
which measurements to include
and how to assign realistic error bars,
given the variable quality
of ground-based data,
the use of heterogeneous instruments
and methods,
and occasional 
misunderstandings about timing systems
and transcription errors
\citep[see, e.g.,][for a careful
examination of several such cases]{Adams+2024}.
Our approach was to
\begin{itemize}

\item include the pre-TESS transit times compiled
by \cite{IvshinaWinn2022}, which are drawn
exclusively from the peer-reviewed literature,

\item exclude the
ground-based data obtained
after the first TESS observation,

\item discard the three data points
with formal uncertainties exceeding 60~s, and

\item impose a minimum uncertainty of 25~s when fitting
the data, to prevent any single data point from
having undue weight.

\end{itemize}

\begin{figure*}
\centering
\includegraphics[width=1.0\textwidth]{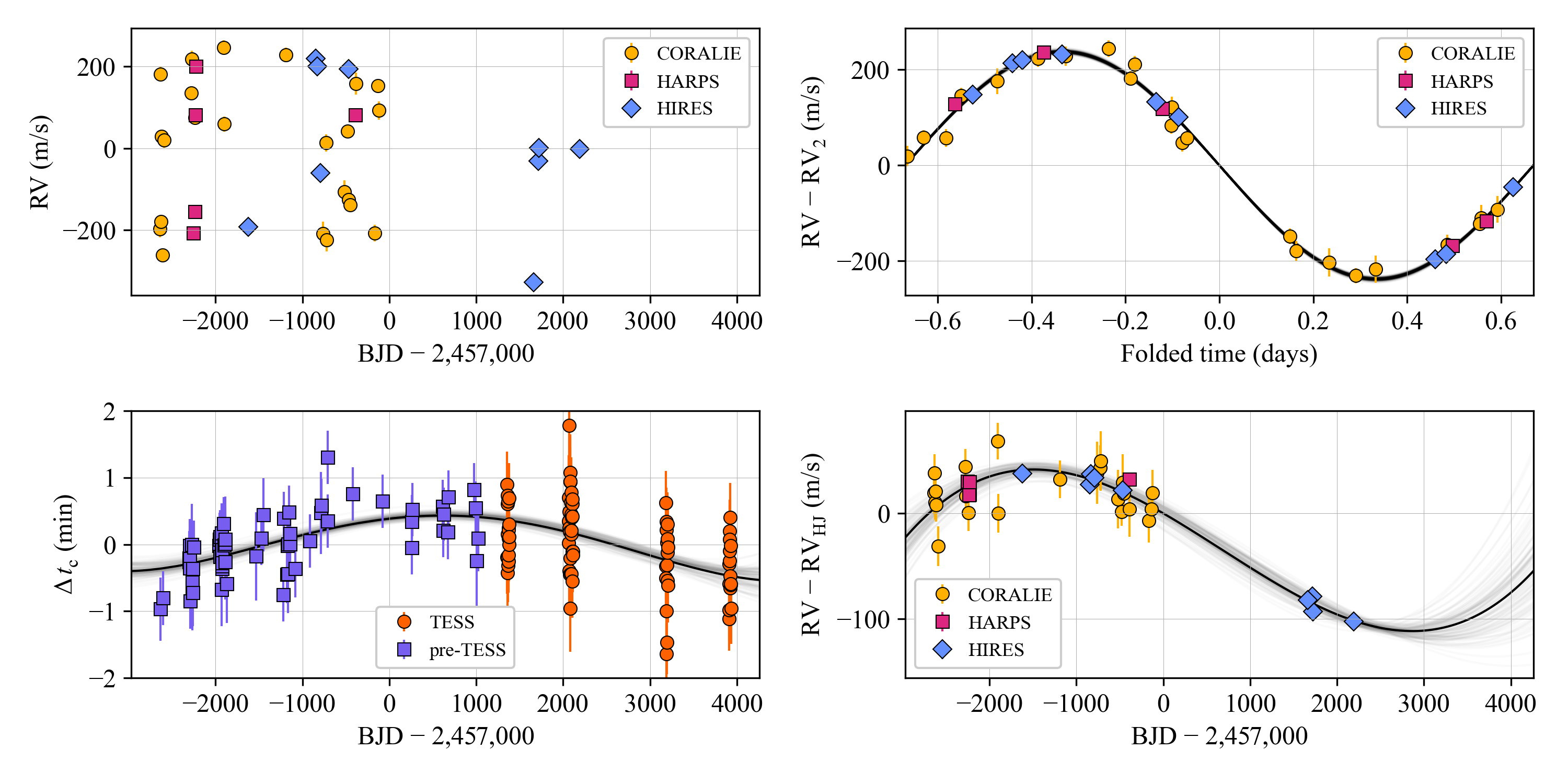}
\caption{{\bf WASP-4.}
{\it Upper left.}---Radial velocities,
Radial velocities, after shifting the data from each spectrograph to have zero mean.
{\it Lower left.}---Deviations between
transit times and the best-fit constant-period
model ($P=1.338231293$~days,
$t_{\rm c} = 2456180.558583$~BJD$_{\rm TDB}$).
In this panel and others,
the black curve is the best-fit
model including the hot Jupiter and
a wide-orbiting companion,
and the gray curves are random draws from
the posterior.
{\it Upper right.}---Radial velocities
as a function of folded time (i.e.\ relative to
the nearest transit time),
after subtracting the best-fit contribution from the
wide-orbiting companion.
{\it Lower right}.---Radial velocities
after subtracting the best-fit contribution
of the hot Jupiter.
\label{fig:wasp4_data}
}
\end{figure*}

After applying these criteria,
the pre-TESS dataset comprises 66 transit times.
The lower left panel of Figure~\ref{fig:wasp4_data}
shows the deviations between the measured transit times
and the best-fit constant-period model, revealing
the downward-curving trend that has been
interpreted as evidence for orbital decay.

\subsection{Timing+RV analysis}

Previous work established that WASP-4\,b's
orbit is nearly circular \citep{Husnoo+2012}, as expected
from tidal interactions, and that
the radial-velocity trend evolves gradually
over the 13-year baseline \citep{Bouma+2020}.
With this in mind, we adopted a model
consisting of a circular orbit for the hot Jupiter,
and a low-order polynomial approximation
for the motion induced by the distant companion:
\begin{align}
\label{eq:wasp4-polynomial-approximation-z1}
\dot{z_1} &= -K_1\sin \left[
\frac{2\pi}{P_1}(t-t_{\rm c,1})
\right]~{\rm and}
\\
\label{eq:wasp4-polynomial-approximation-z2dot}
\dot{z}_2 &= \beta_1 t +
\beta_2 t^2 +
\beta_3 t^3.
\end{align}
The model had nine free parameters:
$K_1$, $P_1$, $t_{\rm c,1}$, the three
$\beta$ parameters,
and the three $\gamma$ offsets (one
per spectrograph).
The best-fit model yielded $\chi^2 = 187.8$
with 172 degrees of freedom,
indicating an acceptable fit.
Posterior sampling
was performed using the
Markov Chain Monte Carlo (MCMC)
algorithm of \citet{GoodmanWeare2010}
as implemented in \texttt{emcee} \citep{ForemanMackey+2013}.
To allow for possible underestimation
of uncertainties, we introduced a velocity jitter
term for each spectrograph.
Table~\ref{tbl:wasp4-parameters} gives the results and 
Figure~\ref{fig:wasp4_data} shows the
corresponding fits.
The model provides a good fit
to all the observables.

\begin{deluxetable*}{cccc}
\label{tbl:wasp4-parameters}
\tablecaption{WASP-4 model parameters}
\tablehead{
  Parameter & Best-fit & Marginalized & Units \\[-0.05in]
            & value    & posterior    & 
}
\startdata
$P_1$ & $1.33823139$ & $1.338231395^{+0.000000023}_{-0.000000024}$ & days \\
$t_{\rm c,1}$ & $-819.441282$ & $-819.441283^{+0.000033}_{-0.000033}$ & BJD$_{\rm TDB} - 2{,}457{,}000$ \\
$K_1$ & $237.6$ & $237.6^{+2.6}_{-2.6}$ & m/s\\
$\beta_1$ & $-17.4$ & $-17.3^{+1.0}_{-1.1}$ & m/s/yr \\
$\beta_2$ & $-1.03$ &  $-1.04^{+0.15}_{-0.14}$ & m/s/yr$^2$\\
$\beta_3$ & $0.180$ & $0.181^{+0.033}_{-0.032}$ & m/s/yr$^3$\\
$\gamma_{\rm CORALIE}$ & $-35.7$ &  $-35.2^{+6.8}_{-6.8}$ & m/s \\
$\gamma_{\rm HARPS}$ & $-64.5$ & $-63.6^{+8.5}_{-9.1}$ &  m/s\\
$\gamma_{\rm HIRES}$ & $-46.9$ & $-46.6^{+3.1}_{-3.4}$ &  m/s\\
$\sigma_{\rm CORALIE}$ & $10.1$ & $10.2^{+6.7}_{-6.3}$ & m/s \\
$\sigma_{\rm HARPS}$ & $11.5$ & $11.6^{+9.6}_{-4.9}$ &  m/s\\
$\sigma_{\rm HIRES}$ & $5.3$ & $5.2^{+2.9}_{-2.0}$ & m/s \\
\hline
$a_2$ & $8.29$  & $7.96^{+0.72}_{-0.32}$ & AU \\
$P_2$ & $25.2$  & $23.7^{+3.3}_{-1.4}$ & yr \\
$K_2$ & $76.9$ & $73.4^{+8.6}_{-3.8}$ & m/s \\
$m_2\sin I_2$ & $7.39$    & $6.93^{+1.10}_{-0.47}$ & $M_{\rm Jup}$
\enddata
\tablecomments{Parameters below the line
were constructed from parameters above the line
assuming $e_2=0$. Data in Column 3 are
based on the 16th, 50th, and 84th percentiles
of the cumulative posterior probability function
for each parameter, marginalized over all other parameters.}
\end{deluxetable*}

\begin{figure}
\includegraphics[width=0.5\textwidth,trim=0.5in 0.5in 0 0]{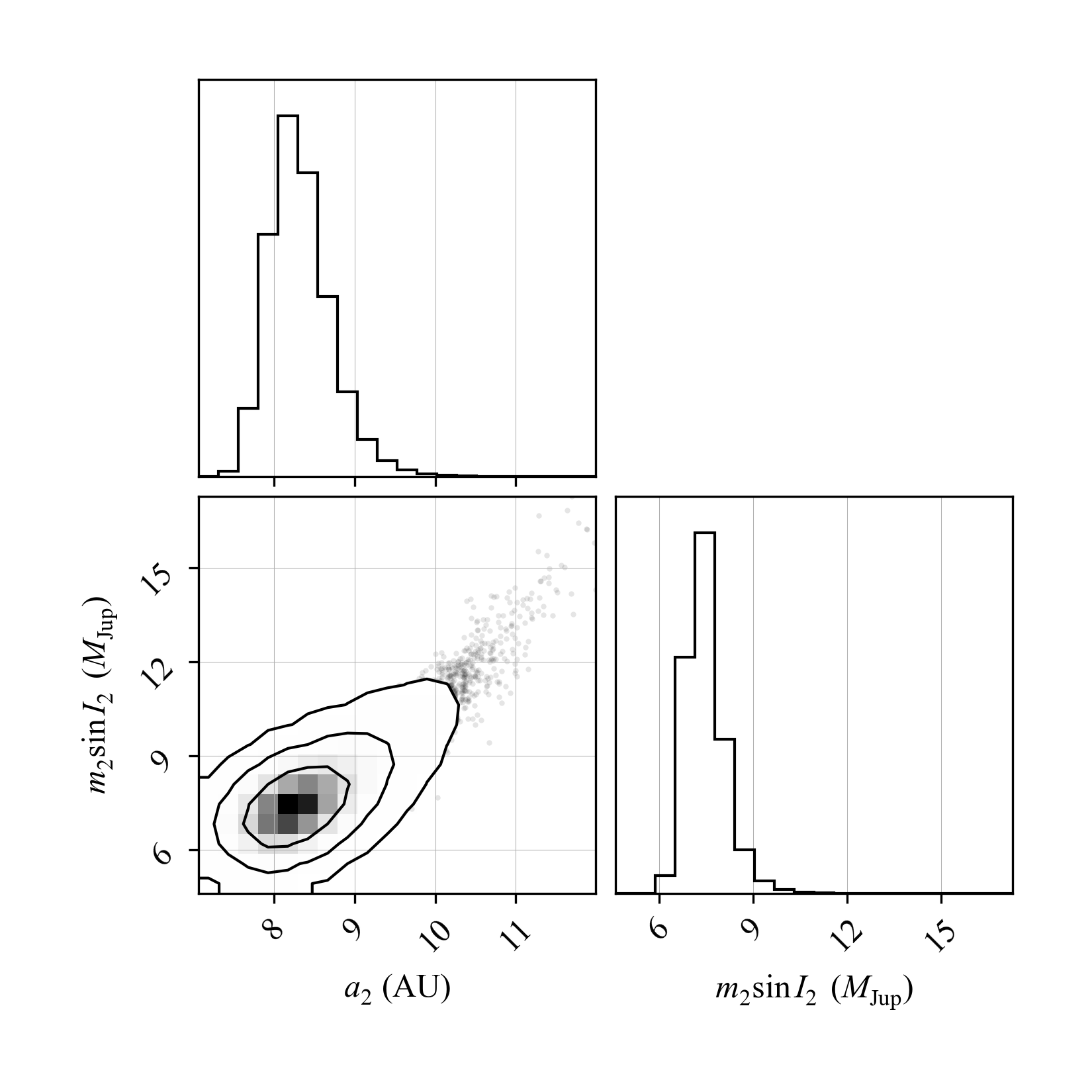}
\caption{{\bf Constraints on the wide-orbiting companion of WASP-4\,b.}
Shown are the posteriors for
minimum mass and orbital separation, assuming the orbit is nearly circular. The contours enclose 68\%, 95\%, and 99.7\% of the probability.
\label{fig:wasp4-companion}
}
\end{figure}

The five parameters of the companion's radial-velocity
orbit ($K_2$, $P_2$, $e_2$, $\omega_2$, and $t_{\rm c,2}$)
cannot be uniquely determined because
the data provide only three constraints ($\beta_1$,
$\beta_2$, and $\beta_3$).
Nevertheless, we can
estimate the companion's
minimum mass and orbital
separation by assuming $e_2\approx 0$
and using
Equations~\ref{eq:beta_values_circular_orbit}--\ref{eq:beta_values_circular_orbit_2} to find
\begin{equation}
n_2 = \sqrt{-\frac{6\beta_3}{\beta_1}}~~{\rm and}~~
K_2 = \frac{1}{n_2} \sqrt{
\beta_1^2 + 
\frac{4\beta_2^2}{n_2^2}
}.
\end{equation}
Adopting a stellar mass of $0.9\,M_\odot$ \citep{Wilson+2008},
we infer a companion with minimum mass
of 6--9~$M_{\rm Jup}$ at 7.5--9~AU
(Table~\ref{tbl:wasp4-parameters}, Figure~\ref{fig:wasp4-companion}).
The posterior distributions
for $a_2$ and $m_2\sin I_2$
have long tails 
extending toward larger masses and
wider separations, but
the radial acceleration of the primary star
is constrained more tightly:
\begin{equation}
\frac{m_2\sin I_2}{a_2^2} = 0.1039^{+0.0084}_{-0.0042}
\frac{ M_{\rm Jup} }{{\rm AU}^2}.
\end{equation}
Allowing the companion to have a nonzero eccentricity would modify
these estimates by factors of order $e_2$.

In summary, a super-Jupiter or brown dwarf companion with an orbital
separation of about 8~AU and a period of about 25 years naturally explains the transit-timing and RV variations of WASP-4, without invoking orbital decay.

\section{WASP-12}
\label{sec:wasp-12}

\begin{figure*}
\centering
\includegraphics[width=1.0\textwidth]{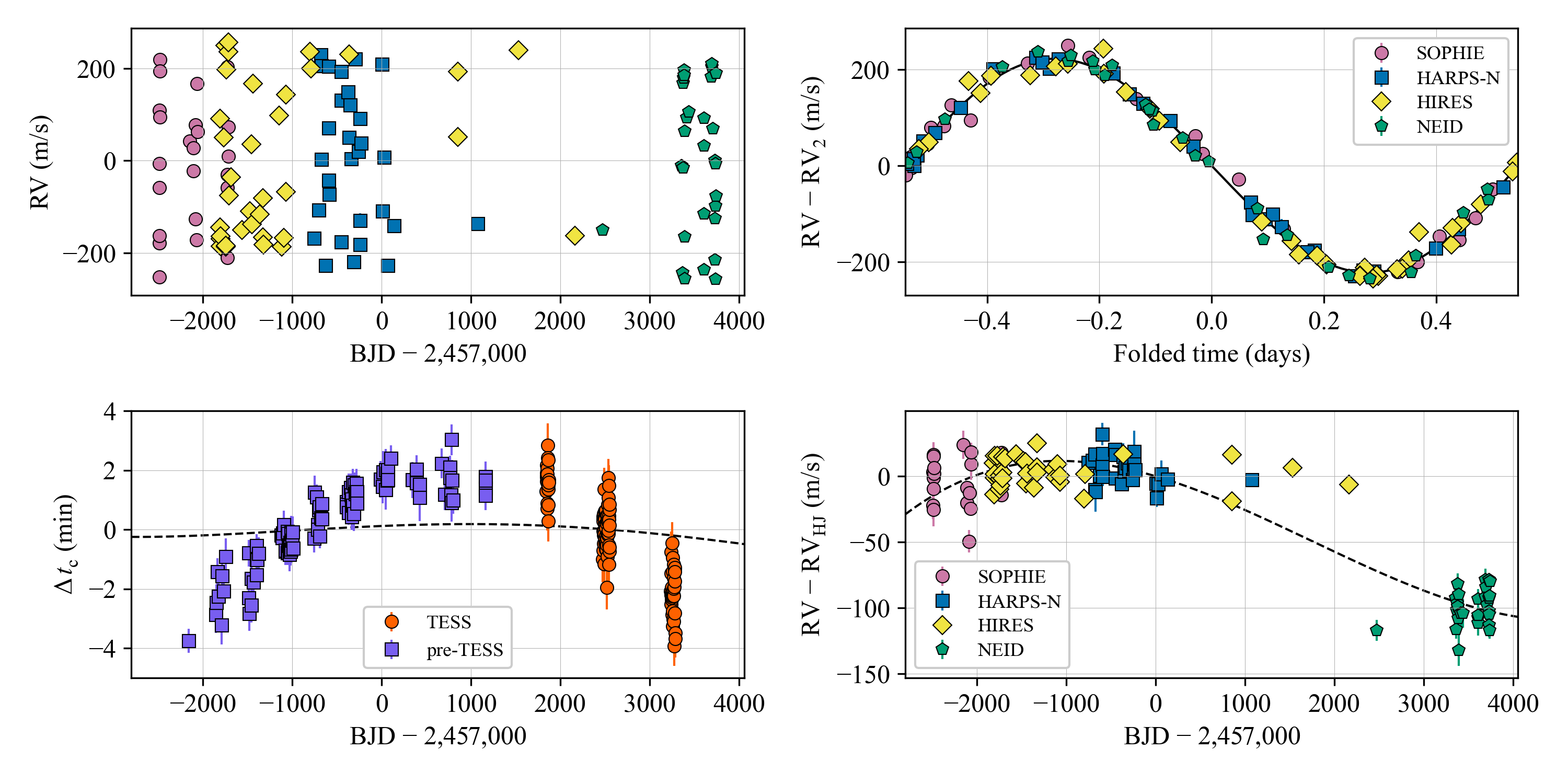}
\caption{{\bf WASP-12.}
{\it Upper left.}---Radial velocities,
after shifting the data from each spectrograph to have zero mean.
{\it Lower left.}---Deviations between transit times
and the best-fit constant-period model ($P=1.091418667$~days,
$t_{\rm c} = 2456305.45552$~BJD$_{\rm TDB}$). In this panel and others,
the black curve is the best-fit
model including the effects of the hot Jupiter and
a wide-orbiting companion,
and the gray curves are random draws from
the posterior.
{\it Upper right.}---Radial velocities
as a function of folded time
(i.e.\ relative to the nearest
transit time),
after subtracting the best-fit contribution of the
wide-orbiting companion.
{\it Lower right}.---Radial velocities
after subtracting the best-fit contribution
of the hot Jupiter.
\label{fig:wasp12_data}
}
\end{figure*}

WASP-12 has been classified
as 6500~K main-sequence star with 
mass 1.4~$M_\odot$ and radius 1.7~$R_\odot$
\citep{Collins+2017, BaileyGoodman2019},
although there are suggestions it may be
somewhat less massive and more evolved
\citep{Weinberg+2017}.
The hot Jupiter WASP-12\,b, discovered
by \cite{Hebb+2009},
has mass 1.5~$M_{\rm Jup}$, radius
$1.9~R_{\rm Jup}$, and period 1.1 days.
With an orbital separation only
three times the stellar
radius, WASP-12\,b lies
precariously close to the Roche limit
\citep{AntonettiGoodman2022}.
We do not attempt to review the vast WASP-12 literature
\citep[see][for an overview]{Haswell2018}.
Our aim was to apply the same joint timing\,+\,RV
analysis used for WASP-4 to test the viability of
the LTTE hypothesis for this system.

TESS data with 120~s sampling
are available from Sectors 20, 43, 44, 45, 71, and 72.
We re-analyzed the SPOC light curves
and combined the measured transit
times with the pre-TESS
compilation of
\cite{IvshinaWinn2022}.
As before,
we excluded ground-based
data obtained after the first TESS observation,
imposed a minimum uncertainty of 25~s
and discarded times with formal
uncertainties exceeding 60~s.

There are three prior sources of radial velocities:
the SOPHIE\footnote{SOPHIE is the Spectrographe pour l’Observation des Phénomènes des Intérieurs stellaires et des Exoplanètes, mounted
on the 1.9\,m telescope at Haute-Provence Observatory, France.} data
of \cite{Hebb+2009},
the HARPS-N\footnote{HARPS-N is the High Accuracy Radial velocity Planet Searcher for the Northern hemisphere, fed by
the 3.6\,m Telescopio Nazionale Galileo
at Roque de los Muchachos Observatory on the La Palma, Canary Islands, Spain.}
data of \cite{Bonomo+2017} and
\cite{Maciejewski+2020},
and the HIRES data of \cite{Yee+2020}.
We supplemented these with 20 new velocities obtained
with the NEID\footnote{NEID stands for
NN-EXPLORE Exoplanet Investigations with Doppler Spectroscopy. It is a stabilized optical spectrograph with resolution
$\approx$120{,}000 \citep{Schwab+2016} mounted
on the WIYN 3.5\,m telescope at Kitt Peak, Arizona.} spectrograph in 2024 and 2025.
The NEID spectra were processed with
version 1.4.2 of the data reduction pipeline.\footnote{https://neid.ipac.caltech.edu/docs/NEID-DRP/}
Spectra obtained during transits were excluded,
and the time series was thinned
to enforce a five-day minimum time spacing.
The resulting velocities are given in 
the Appendix (Table~\ref{tbl:wasp12-rv}).
To account for stellar activity, we adopted a velocity jitter of 10~m/s,
the approximate value determined in
single-planet fits over time intervals
less than a few years \citep{Bonomo+2017,Maciejewski+2020, Yee+2020}.

Figure~\ref{fig:wasp12_data} shows
the RVs and transit timing deviations.
The 400~m/s RV variations
are due to the hot Jupiter, with no obvious
long-term trend.
The transit timings, however,
display a clear
quadratic trend.

To test the LTTE hypothesis,
we used a model similar to that used
for WASP-4,
modified
only to account
for the RV anomalies caused
by the star's tidal distortion \citep{Arras+2012}.
For small distortions,
the effect introduces
a sinusoidal signal
with a period of $P_1/2$, an amplitude of a few m/s, and a  
phase shift of 90$^\circ$ relative to the main signal.
We modeled this effect by the simple
expedient
of assigning the planet a fictitious eccentricity
and fixing $\omega=-\pi/2$
\citep[for more details, see][]{Arras+2012,Maciejewski+2020}.

The best-fit model is shown in
Figure~\ref{fig:wasp12_data} (dashed black curves).
The fit is very poor,
with $\chi^2 = 1{,}531$ and 323 degrees of freedom.
The model attempts to reproduce the quadratic
timing trend with a suitably strong
line-of-sight acceleration (lower left)
but is stymied by the constraint
imposed by contemporaneous RV data
(lower right), especially
the HIRES data (yellow diamonds).

In short, the joint timing\,+\,RV analysis
firmly rules out LTTE as the explanation
for the shrinking transit period of WASP-12\,b.
This result corroborates earlier studies and
provides a useful contrast with WASP-4\,b,
where LTTE offers a compelling explanation.

\medskip

\section{Kepler-1658}
\label{sec:kepler-1658}

\begin{figure*}
\centering
\includegraphics[width=1.0\textwidth]{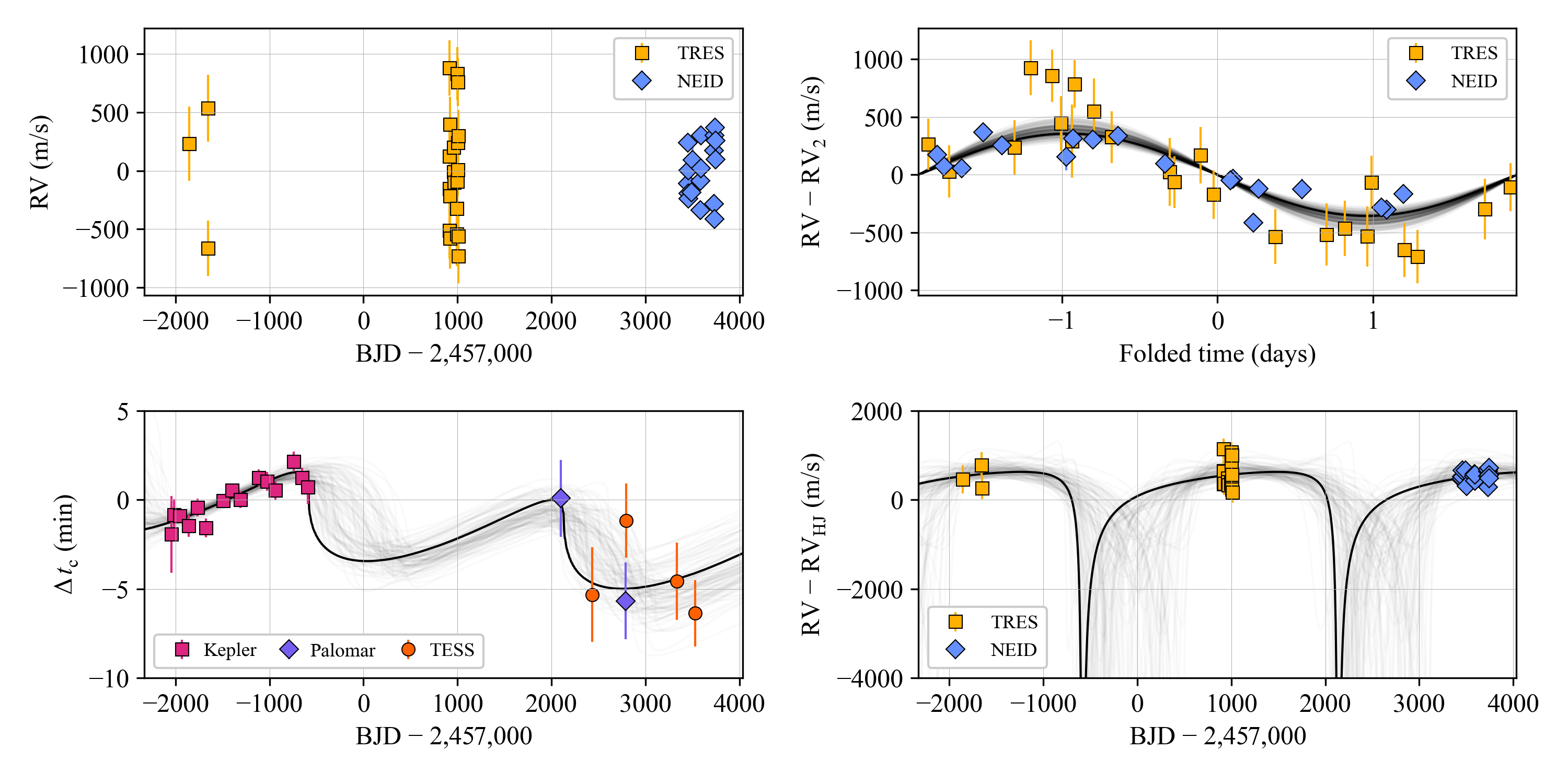}
\caption{{\bf Kepler-1658.}
{\it Upper left.}---Radial velocities,
after shifting the data from each spectrograph to have zero mean.
{\it Lower left.}---Deviations between
transit times and the best-fit
constant-period model ($P=3.849363145$~days,
$t_{\rm c} = 2455005.925555$~BJD$_{\rm TDB}$). 
In this panel and others,
the black curve is the best-fit
model including the effects of the hot Jupiter and
a wide-orbiting companion,
and the gray curves are random draws from
the posterior.
{\it Upper right.}---Radial velocities
as a function of folded time (i.e.\ relative to
the nearest transit time)
after subtracting the best-fit contribution of the
wide-orbiting companion.
{\it Lower right}.---Radial velocities
after subtracting the best-fit contribution
of the hot Jupiter.
\label{fig:kepler-1658_data}
}
\end{figure*}

Kepler-1658 was first described in detail by \cite{Chontos+2019},
using data from NASA's Kepler mission \citep{Borucki+2010}
and follow-up observations.
Early in the mission, a sequence of transits was detected
with a period of 3.9 days and a depth of 0.13\%.
The source was inconsistently classified
in mission catalogs,
appearing as a planet candidate in
some and a false positive in others.
Confirmation of Kepler-1658\,b as a planet was based mainly
on the detection of radial-velocity variations
with the expected period and phase, an amplitude
consistent with a planetary mass
($\approx$\,6~$M_{\rm Jup}$), and no
evidence for line-bisector variations.
The host star is a subgiant with
effective temperature 6200\,K, mass
$1.4\,M_\odot$, and radius $2.9\,R_{\odot}$.
Because tidal dissipation
rates are expected to increase
as a star evolves into the subgiant phase,
the system was highlighted as a
favorable target for detecting orbital
decay.

\cite{Vissapragada+2022} analyzed TESS data from
Sectors 41, 54, and 55, as well as
two ground-based observations
with the Wide Field Infrared Camera
(WIRC) on the Hale 5\,m telescope
at Palomar Observatory, California.
When combined with the Kepler data,
these transit times
revealed apparent orbital decay 
at a rate of 130~ms/yr,
five times faster than for WASP-12\,b
and greatly
exceeding theoretical expectations
\citep{Barker+2024}.

Could the decreasing transit period be caused by the light
travel-time effect of an outer companion?
\cite{Vissapragada+2022} argued that the existing
RV data
were sufficient to
rule out this possibility.
Our joint timing\,+\,RV analysis,
described below,
shows that this conclusion was premature.
More fundamentally, our analysis of new
data and re-analysis of older data
point to a different interpretation
of Kepler-1658 altogether:
it is not a hot Jupiter system,
but rather an unresolved
triple star system.

\subsection{Radial Velocities}

\cite{Chontos+2019} used TRES\footnote{TRES is the Tillinghast Reflector Echelle Spectrograph,
mounted on the 1.5\,m telescope at Mt.\ Hopkins, Arizona.} to measure
23 RVs with typical
uncertainties of 150~m/s.
Three measurements
were made in 2009 and 2010,
and the other 20 came after a seven-year gap.
To this dataset we added 22 velocities from NEID
obtained in 2024--2025.
The official NEID pipeline reduction (v1.4.3)
produced velocities with anomalously large uncertainties
and no sign of the known 3.9-day periodicity.\footnote{In retrospect, this might have been
due to the spectral anomalies
discussed in Section~\ref{subsec:kepler-1658-problems}.}
We therefore performed a custom extraction
using the SERVAL template-matching code \citep{zechmeister2018},
as adapted for NEID by \cite{stefansson2022}.
The resulting velocities had uncertainties of 50~m/s, more
in line with expectations.

The RV data are shown in the
upper left panel of Figure~\ref{fig:kepler-1658_data}.
The peak-to-peak
variations of about 1{,}000~m/s have been attributed
to the hot Jupiter.
Beyond this, as noted by \cite{Vissapragada+2022},
there is no evidence for a long-term trend.
The constraint is weaker
than it may appear, though, because
the TRES and NEID data can be shifted
relative to each other by an arbitrary constant.
The primary leverage on any RV trend
comes from the consistency between the earliest three
TRES velocities and those obtained
seven years later.

\subsection{Transit Times}

We re-analyzed the TESS light curves from Sectors
41, 54, and 55, and incorporated
new TESS data
from Sectors 74, 75, 81, and 82.
Because single transits are barely detectable in the TESS photometry,
we abandoned attempts to measure individual transit times.
Instead, we grouped the data into four multi-transit blocks:
Sector 41, Sectors 54 \& 55,
Sectors 74 \& 75, and Sectors 81 \& 82.
Within each block, the light curve was fitted
using the same basic method as \cite{IvshinaWinn2022},
but assuming strict periodicity, with the
key parameter being the
transit time of the event closest
to the block center.
There was no requirement for strict periodicity
across different blocks.
This approach
offers greater
robustness
at the expense of sensitivity to short-term
variations. 
\cite{Vissapragada+2022} used
a similar procedure to
analyze the Kepler data,
treating the observing
quarters as independent blocks.
We adopted without modification
their reported transit times from Kepler and
from WIRC.

The timing deviations are displayed in the
lower left panel of Figure~\ref{fig:kepler-1658_data}.
Even without reference to the fitted curves,
the data indicate that the average transit
period during the Kepler mission
was longer than during 
the TESS era. This was the basis
of the claim of orbital decay.

\subsection{Analysis}
\label{subsec:kepler-1658-analysis}

The case of Kepler-1658 differs
from that of WASP-4 in two important
respects: no long-term RV trend is evident,
and the datasets
are relatively sparse and irregularly spaced.
The long observational gaps
leave open the possibility
of variations
that are not well
approximated by low-order polynomial functions.
For this reason,
we modeled
the effects of the hypothetical companion
with a full Keplerian orbit parameterized by
$\{K_2, P_2, t_{\rm c,2}, e_2, \omega_2\}$,
resulting in 10 free parameters
and 62 data points.

We scanned $10^4$ trial values of $P_2$,
log-uniformly distributed between $7$~and $10^5$~days.
For each choice of $P_2$,
we optimized the other parameters
and recorded the value of $\chi^2$.
The optimal solution was found at
$P_2 \approx 2500$~days,
with $\chi^2= 187$ with
52 degrees of freedom --- a poor fit.
We proceeded under
the tentative assumption
that the RV uncertainties
had been underestimated, which
seemed plausible since
precise RV extraction codes are not optimized for
broad-lined stars such as
Kepler-1658
($v\sin i\approx 34$~km/s).
By adding jitter terms 
of $\sigma_{\rm TRES}=180$~m/s
and $\sigma_{\rm NEID}=100$~m/s
in quadrature with the formal
uncertainties,
the $\chi^2$ of the
best-fit model was lowered
to equal the number of degrees of freedom.

In the best-fit model,
the companion is a $\approx$0.8$\,M_\odot$ star
that completes a full orbit
between the first and second
batches of TRES observations,
and another orbit during the seven-year interval
separating the TRES
and NEID observations.
Figure~\ref{fig:kepler-1658_data}
shows the model curves.
The next-best model has $P_2 \approx 8{,}300$ days ($\chi^2=77$),
allowing the companion
to execute half an orbit
during each gap.
The third best model
has $P_2 \approx 1{,}970$~days ($\chi^2=81)$
in which each gap accommodates 1.5 orbits.

While this exercise demonstrated that the LTTE
hypothesis cannot be ruled out as the explanation
for the observed change in the transit period, the
models are unconvincing. The required
jitter terms are large, and the models seem
to ``cheat'' by arranging for the strongest
variations to occur during gaps
in the dataset.
This prompted us to re-examine the
Kepler-1658 data more closely.

\subsection{Problems with the planet hypothesis}
\label{subsec:kepler-1658-problems}

We began by checking for consistency of the RV semi-amplitudes
measured by the two different instruments.
Fitting each dataset with a sinusoidal function
having the known period and phase\footnote{When fitting the TRES
data, we did not include the three 
earliest data points from 2009,
to reduce the possible effects of any long-term trend.}
yielded
$K_{\rm TRES} = 561\pm 75$~m/s and
$K_{\rm NEID} = 318\pm 40$~m/s,
a significant difference of $\Delta K = 243\pm 85$~m/s.
The discrepancy can be seen
in Figure~\ref{fig:kepler-1658_data},
where the blue and orange points trace different amplitudes.

\begin{figure*}
\centering
\includegraphics[width=1.0\textwidth]{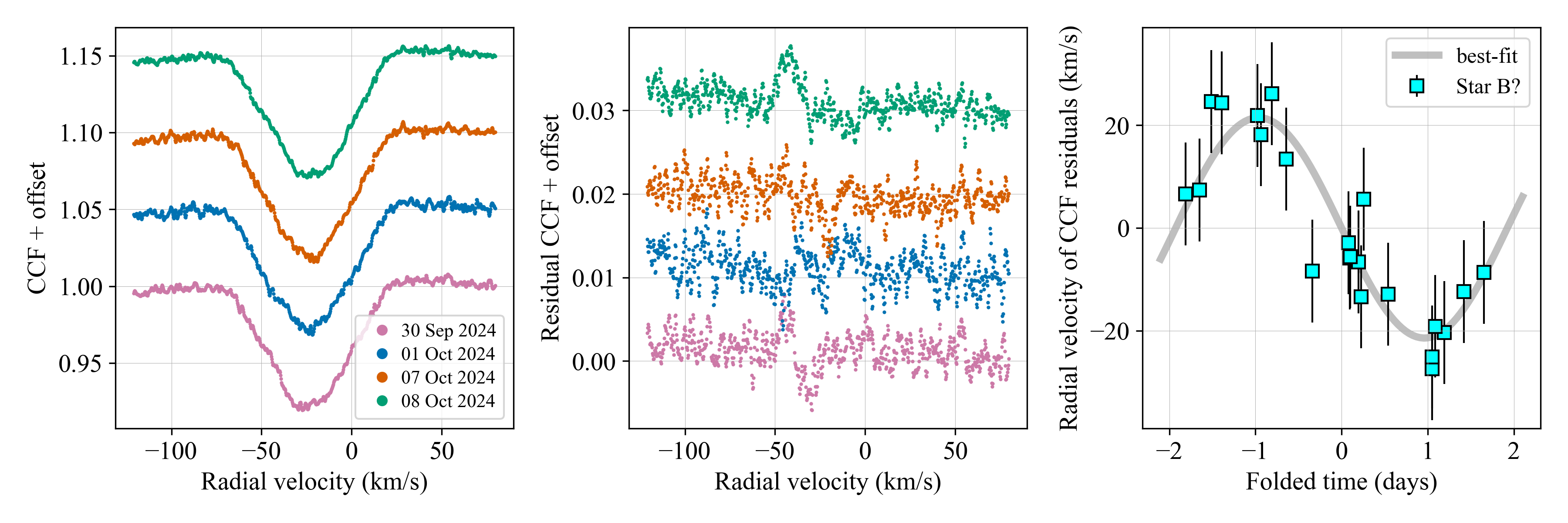}
\caption{{\bf Spectral line variations
of Kepler-1658.}
{\it Left.}---Representative cross-correlation
functions (CCFs) based on NEID spectra.
They are centered on the systemic velocity
of $-24$~km/s and have widths of about
30~km/s due to stellar rotation.
There are also night-to-night shape
variations.
{\it Middle.}---Isolation of CCF variations.
The median NEID CCF was
subtracted from individual CCFs,
revealing anomalies of amplitude $\sim$$10^{-2}$
that shift in velocity.
{\it Right.}---Velocity shift of the pattern
of residuals relative to an arbitrarily chosen
night, as a function of time folded
with the 3.8-day transit period.
The variation is consistent with 20~km/s
orbital motion of an unresolved binary
with the expected period and phase.
\label{fig:kepler-1658_ccfs}
}
\end{figure*}

We then inspected the NEID cross-correlation functions (CCFs),
obtained by correlating each spectrum against
an F-star template.
The CCFs exhibit night-to-night variations at the percent level
(Figure~\ref{fig:kepler-1658_ccfs}).
This brought to mind
a case presented by \cite{Mandushev+2005},
in which a rapidly rotating F star
was initially thought to have a transiting 
brown dwarf, but the signal was eventually shown
to arise from the blended light of an F star and a fainter
eclipsing binary. The shifting lines of the binary
produced subtle distortions within the F star's
broad lines that were mistaken for RV variations.

Suspecting that Kepler-1658 could be a Mandushev-type
false positive, we looked for evidence of orbital
motion in the shifting pattern of CCF residuals.
We subtracted the median NEID CCF
from each individual CCF, and then cross-correlated
the residuals.
The resulting velocity shifts,
shown in the right panel of
Figure~\ref{fig:kepler-1658_ccfs} follow a sinusoid
with the same period and phase as the transit signal,
and a semi-amplitude of about 20 km/s.
This is precisely the signature expected of a blended
eclipsing binary masquerading as a planetary companion.

\begin{figure*}
\includegraphics[width=1.0\textwidth]{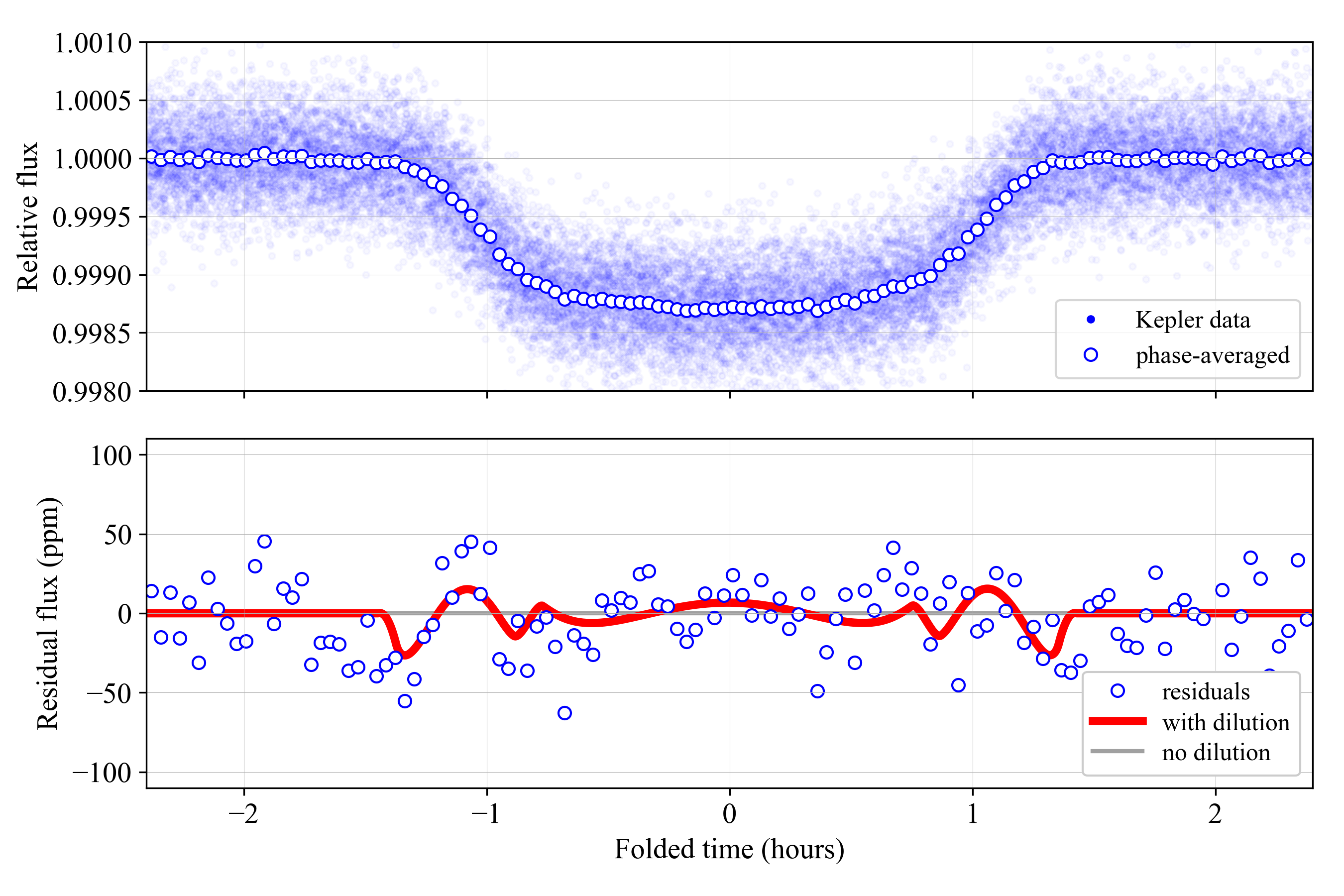}
\caption{{\bf Evidence for dilution in the
light curve of Kepler-1658.}
{\it Top.}---Kepler light curve,
with its native 1-minute sampling
as well as the phase-averaged version
used for model fitting.
{\it Bottom.}---Residuals between the data
and the best-fit transiting-planet
model. The pattern
of residuals is well-fitted by a model
in which the ``transits'' are eclipses
of a faint binary that contributes
only 1.7\% of the total light.
\label{fig:kepler-1658_Kepler_Light_Curve}
}
\end{figure*}

Seeking corroborating evidence, we re-examined
the Kepler photometry.
A median light curve was constructed from
all the 1-min data: the data were phase-folded using the transit
period, and phase-averaged to 2.3-min cadence (124 data points) to speed up further analysis.
The standard deviation 
outside of transits was 22~parts
per million (ppm), which we adopted as
the per-point uncertainty.
Fitting the light curve with the
\cite{MandelAgol2002} transit model
gave $\chi^2 = 134.5$
with 117 degrees of freedom.
While this is statistically 
acceptable, there are 
structured patterns in the residuals,
especially during ingress and egress
(Figure~\ref{fig:kepler-1658_Kepler_Light_Curve}).

Motivated by the blended binary hypothesis,
we fitted the light curve with a diluted model:
\begin{equation}
F(t) = (1-\epsilon) + \epsilon\,\delta(t),
\end{equation}
where $\delta(t)$ is the \cite{MandelAgol2002}
model and $\epsilon$ is the fractional light
contribution of the binary.
Adding the single parameter $\epsilon$
lowered $\chi^2$ by 20.3 units and
reduced the structure in the residuals,
yielding
$\epsilon=0.0173^{+0.0020}_{-0.0014}$.

\begin{figure}
\includegraphics[width=0.45\textwidth]{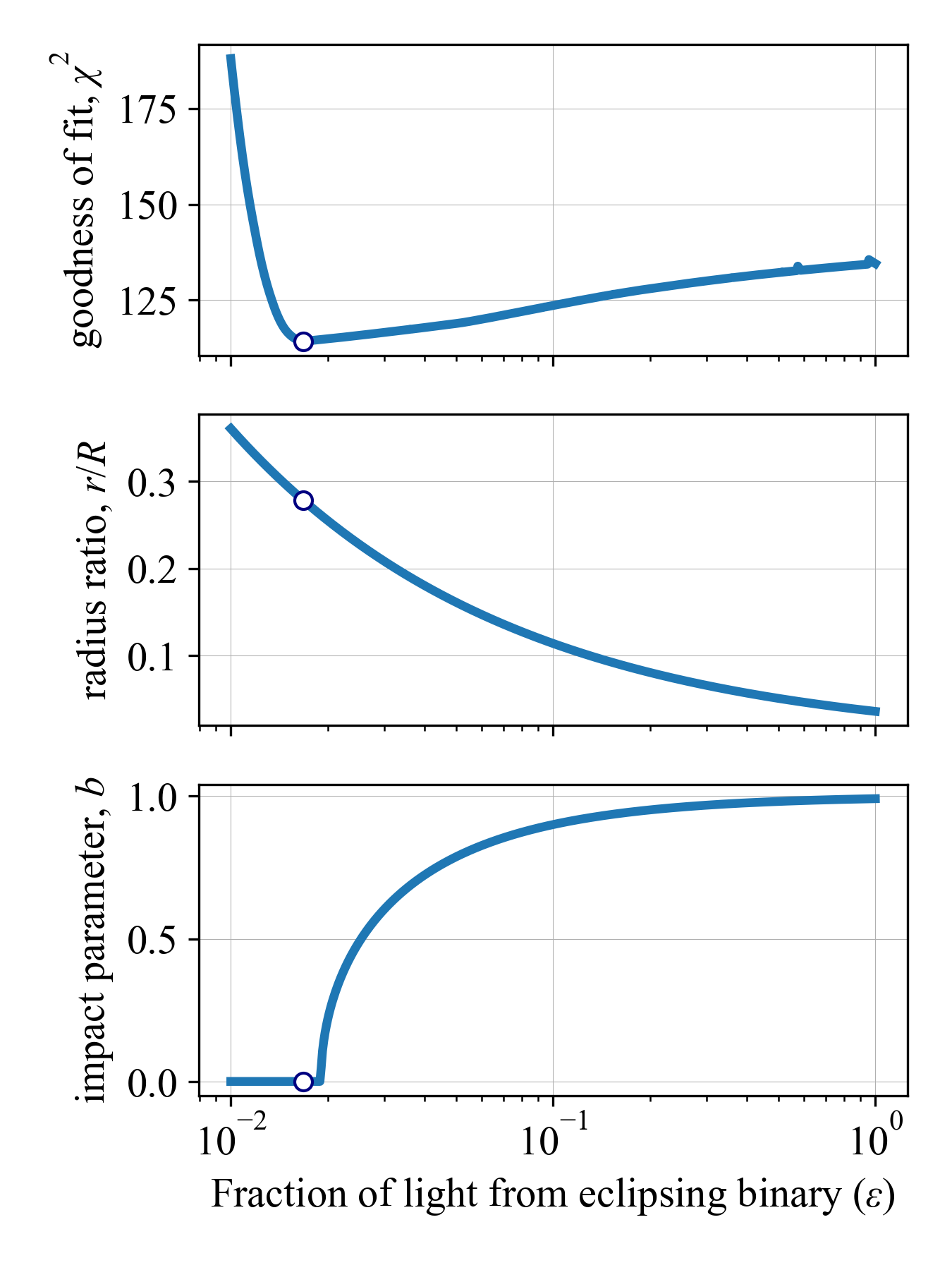}
\caption{{\bf Constraints on dilution from the Kepler-1658 light curve.}
{\it Top.}---Goodness of fit, $\chi^2$, as a
function of $\epsilon$, the fraction of the
light contributed by the blended binary.
{\it Middle and bottom.}---Corresponding
values of the radius ratio between the two stars
of the eclipsing
binary, and the eclipse impact parameter.
The data favor $\epsilon\approx 0.017$,
$r/R\approx 0.27$,
and $b\approx 0$.
\label{fig:kepler-1658_Chi2_vs_epsilon}
}
\end{figure}

Figure~\ref{fig:kepler-1658_Chi2_vs_epsilon} shows
that smaller and larger values of $\epsilon$ are disfavored.
When $\epsilon$ is increased from the best-fit
value of 0.0173, the model preserves the transit depth 
by reducing the radius ratio $r/R$,
and preserves the transit duration by increasing
the impact parameter $b$, but the fit to the subtle
details of ingress and egress
becomes worse.
Lowering $\epsilon$ below the best-fit value
fails because the
transit duration cannot be preserved;
the impact parameter
is already at its minimum value.

In summary, the evidence against the planetary hypothesis
for Kepler-1658\,b seems compelling: inconsistent RV
amplitudes between instruments,
sinusoidal CCF variations with the same period
and phase as the transits,
and evidence of strong dilution in the Kepler light curve.
Taken together, these findings point to a hierarchical
triple system rather than a hot Jupiter.

\subsection{Hierarchical triple-star model}

Our next task was to identify plausible
parameters for a hierarchical triple
that reproduces all the relevant observations.
We denote the broad-lined F star as
component A, and the members of the 3.8-day eclipsing
binary as B and C.
The goal was to determine $M_{\rm B}$, $M_{\rm C}$,
and the separation of A from the B/C pair, subject
to five observational constraints:
\begin{itemize}
\item the apparent transit and occultation depths observed
by Kepler,
\item the $J$-band apparent transit depth seen
in the WIRC light curve,
\item the dilution factor of $\epsilon \sim 10^{-2}$
inferred from the Kepler light curve,
\item the $\sim$20~km/s sky-projected
orbital speed indicated by the CCF residuals,
and
\item the apparent decline of the eclipse period
by 130~ms/yr.
\end{itemize}
To relate the stellar parameters to these observables,
we assumed B and C
are main-sequence dwarfs
and employed
the ``Mamajek'' zero-age main-sequence relations\footnote{Table 6 of 
\citet{PecautMamajek2013}, as updated online at
\url{https://www.pas.rochester.edu/~emamajek/EEM_dwarf_UBVIJHK_colors_Teff.txt}}
linking mass, radius, and
absolute magnitudes in various
bandpasses. We also needed transformations into Kepler magnitudes,
which we obtained from
\cite{Brown+2011}.

The depth of the secondary eclipse specifies the Kepler-band
flux ratio between C and A:
\begin{equation}
\delta_{\rm occ} =
\frac{F_{\rm C}}{F_{\rm A} + F_{\rm B} + F_{\rm C}} \approx
\frac{F_{\rm C}}{F_{\rm A}}.
\end{equation}
Since A's properties are observed directly,
we can use the flux ratio to calculate
C's Kepler-band absolute magnitude and the
corresponding stellar mass.
To match the observed depth
of 61~ppm, star C should be an M dwarf
with $M_{\rm Kep} = 12.47$,
$M_J = 8.67$,
mass $0.19\,M_\odot$, and 
radius $0.23\,R_\odot$.

The predicted rate of change of the period due to LTTE is
\begin{equation}
\frac{\delta P}{P}
\sim
\frac{P}{c}\,\frac{GM_{\rm A}}{a^2}
\sim
130~{\rm ms/yr}
%\left( \frac{P}{3.8~{\rm d}} \right)
\left( \frac{M_{\rm A}}{1.4\,M_\odot} \right)
\left( \frac{a}{50~{\rm AU}} \right)^{\!-2}\!\!\!.
\end{equation}
Consistency with the observed rate
is achieved if the eclipsing binary
is located $a \sim 50$~AU from the F star.
At the Gaia-determined distance
of 788~parsecs 
\citep{Gaia+2023}\footnote{\url{https://gaia.ari.uni-heidelberg.de/singlesource.html}},
the implied angular separation
is $0.06$~arcsec, consistent with
being unresolved.

\begin{figure}
\includegraphics[width=0.45\textwidth,trim=0.5in 0 0 0]{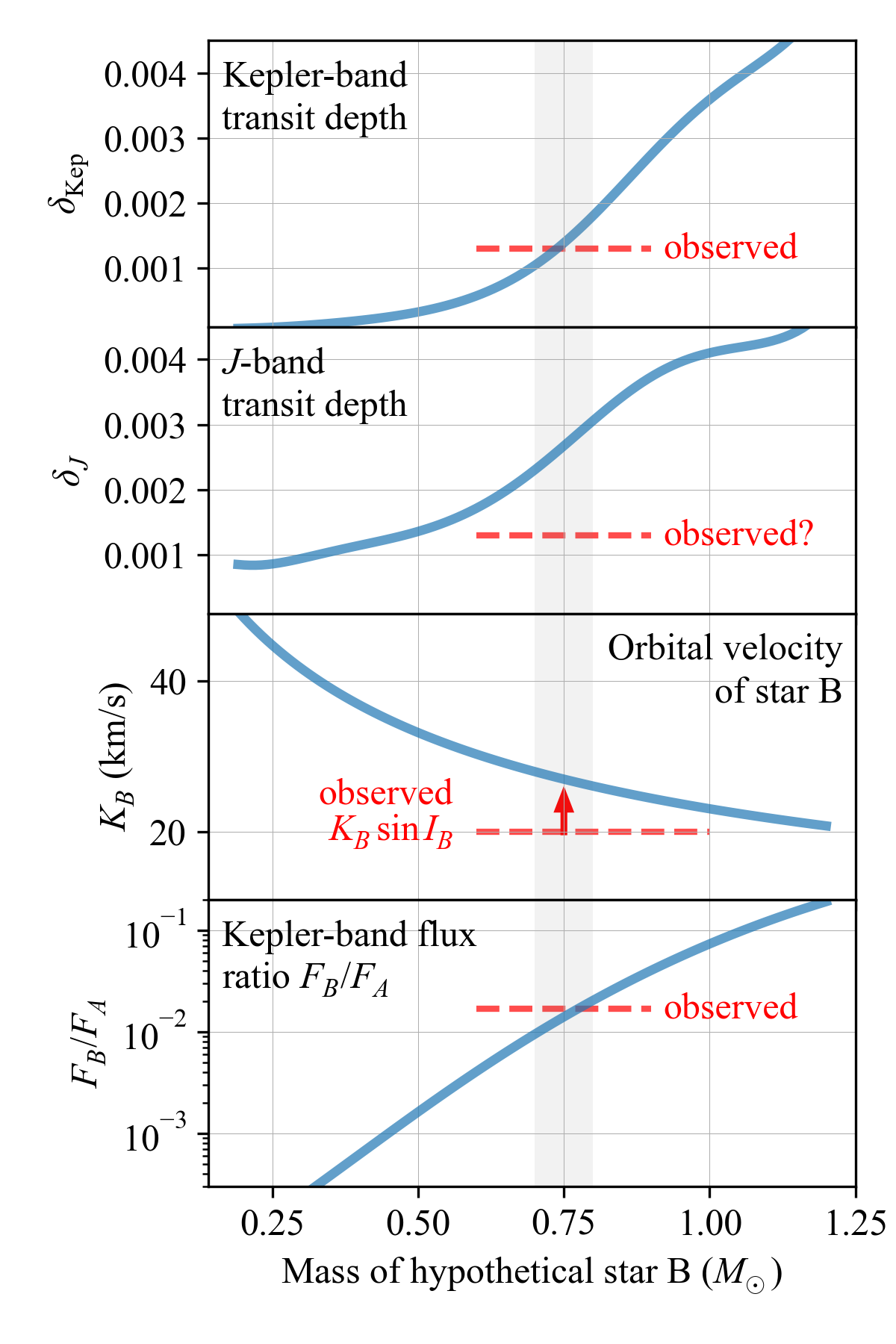}
\caption{{\bf Kepler-1658 as a hierarchical triple star.}
The dependence of various observables is
shown as a function of the mass
of star B. Choosing $M_B\approx 0.75\,M_\odot$
(light gray band)
is compatible with all the observables
except the $J$-band apparent transit depth
(see the text for details).
\label{fig:kepler-1658-FP-scenario}
}
\end{figure}

The apparent transit depth is
\begin{equation}
\delta_{\rm tra} \approx \frac{F_{\rm B} \left( \frac{R_{\rm C}}{R_{\rm B}} \right)^{\!2}} {F_{\rm A} + F_{\rm B} + F_{\rm C}},
\end{equation}
which can be considered a function of the single parameter $M_{\rm B}$
using the Mamajek relations.
Figure~\ref{fig:kepler-1658-FP-scenario}
shows that $M_{\rm B}\approx 0.75\,M_\odot$
(a K dwarf) simultaneously satisfies
the constraints
on the Kepler-band
transit depth (top)
and the 1.7\% flux
contribution of the binary
inferred from the diluted model of the
Kepler light curve (bottom).
With star B at $0.75\,M_\odot$,
the orbital velocity is predicted to be 27~km/s,
consistent with the sky-projected velocity
of 20~km/s inferred from the CCF residuals.

The only tension is
with the $J$-band transit depth,
which is predicted to be almost twice
as large as displayed
in Figure~2 of \cite{Vissapragada+2022}.
It should be noted, though, that the $J$-band
light curve was constructed with a flexible
model for systematics and
a prior constraint enforcing the transit depth to be similar
to the Kepler-band value.\footnote{According
to \cite{Vissapragada+2022},
the time series of WIRC aperture fluxes
was fitted with a model having free parameters
for the weights of 9-10 comparison
stars and the strengths
of correlations with instrumental
parameters such as focus and centroid
position.} It seems plausible
that allowing the $J$-band transit
depth to vary freely would yield a broader
range of values consistent with our prediction.

The triple-star model
also provides natural explanations for a few
peculiarities of the planet hypothesis:
\begin{itemize}

\item For a transiting planet,
the bulk density of the star
can be calculated as \citep{SeagerMallen-Ornelas2003, Winn2010}
\begin{equation}
\rho_\star = \frac{3\pi}{GP^2}\left(\frac{a}{R} \right)^{\!3}.
\end{equation}
Fitting the Kepler light curve with $\epsilon=0$
gives $\rho_\star = 0.19\pm 0.04$~g/cm$^3$,
which disagrees by 4.5-$\sigma$
with the value of $0.083\pm 0.008$~g/cm$^3$
determined by \cite{Chontos+2019} using the independent
technique of asteroseismology.\footnote{
By allowing the hot Jupiter's orbit
to be eccentric, \cite{Chontos+2019} relieved but did not eliminate this tension; in their joint model
the median of the posterior for $\rho_\star$ differs
by 3-$\sigma$ from the asteroseismic value (cf.\ their
Tables 2 and 4).}
In the triple-star model, the eclipse parameters
are unrelated to the bulk density of star A
and there is no tension.

\item If interpreted as reflected light from the hot
Jupiter, the secondary eclipse depth
of 62~ppm
implies an albedo of $0.724_{-0.081}^{+0.090}$. While this
is physically possible, most other
albedo measurements are $\lesssim 0.3$ (see, e.g., Figure 10 of \citealt{Wong+2021}).
In the triple-star model, there is nothing
unusual about the secondary eclipse depth.

\item The Kepler light curve shows
0.1\% variations with a period
of 5.7 days.
For this to be due to rotation of the F star, 
it is necessary to invoke high-latitude starspots
and 20-40\% latitudinal differential rotation
\citep{Chontos+2019}.
In the triple-star model, the modulation can be
attributed to star B.

\end{itemize}

\section{Discussion}
\label{sec:discussion}

Long-term transit timing of hot Jupiters offers many
possible rewards:
the detection of orbital decay,
the discovery of additional orbiting bodies \citep[see, e.g.,][]{Korth+2024,Yang+2025},
and possibly even the effects of the planet's
own tidal deformation \citep{RagozzineWolf2009}.
Another attraction of transit timing is that
relatively small telescopes -- even amateur telescopes -- can make
scientifically valuable contributions.
The arrival of high-quality
all-sky data from the TESS mission has improved the reliability
and expanded the accessibility of transit-timing studies.
However, the cases examined in this paper
underscore that transit timing by itself is usually not enough to earn
the scientific rewards. Light travel-time effects
from outer companions can mimic orbital decay,
and long-term RV monitoring is essential
to distinguish these possibilities.

A more specific lesson from the work described in this paper
is that measuring or placing an upper bound on a
linear RV trend is not sufficient to rule out the LTTE
hypothesis.
For WASP-4, this was because the RV trend was
not well described as linear over the relevant time span.
For datasets with sparse coverage,
similar to that of Kepler-1658,
large gaps admit the possibility of shorter-term variations that
are poorly described by linear trends.

Finally, the Kepler-1658 story is a reminder
that false positives remain a real hazard, even
after two decades of experience with wide-field
transit surveys. Blended eclipsing binaries
bound to bright primaries, especially when the primaries
are rapid rotators, can evade standard vetting procedures
and mimic planetary signals to high precision.
Careful scrutiny of spectral line shapes and
light-curve details is essential.

Of the three systems considered here, only WASP-12\,b
shows persuasive evidence for orbital decay,
supported not only by joint timing\,+\,RV analyses but
also the observation
that the occultation period is
shrinking in step with the transit period \citep{Yee+2020}.
As the timing baseline from TESS grows longer,
and after the PLATO and Roman missions commence
\citep{Rauer+2025, Carden+2025}
additional candidates for orbital decay are likely
to emerge. 
Interpreting them
will require persistence, long-term RV monitoring,
and careful vetting to distinguish
genuine orbital changes from
confounding effects.

% \medskip
% \medskip
% \medskip
% \medskip

{\it Postscript.}---After this manuscript
was submitted and posted on the arXiv, several colleagues provided further evidence supporting the false-positive interpretation of Kepler-1658\,b. S.\ Vissapragada found that the $J$-band WIRC light curve is consistent with the $\approx$0.3\% transit depth predicted in Section~\ref{sec:kepler-1658}. S.\ Mahadevan and C.\ Cañas pointed out that \cite{Fleming+2015} had previously reported a discrepancy between the RV amplitudes measured at optical and near-infrared
wavelengths.

\section{Acknowledgments} 
\label{sec:acknowledgments}

We thank Shreyas Vissapragada, Dan Huber,
Özgür Baştürk, Suvrath Mahadevan, and Caleb Cañas
for feedback on the manuscript.
J.N.W.\ gratefully acknowledges support from the NASA TESS project and a NASA Keck PI award.
The work described in this
paper was based in part on observations with the NEID
instrument on the WIYN 3.5\,m telescope
at Kitt Peak National
Observatory. Kitt Peak is
a facility of NSF's NOIRLab, managed by the Association
of Universities for Research in Astronomy (AURA).
The WIYN telescope is a joint facility of NOIRLab, Indiana University, the University of Wisconsin-Madison, Pennsylvania State University, Purdue University, and Princeton University.
NEID was funded by the NASA-NSF Exoplanet Observational
Research (NN-EXPLORE) partnership and built by Pennsylvania State University. The NEID
archive is operated by the NASA Exoplanet Science Institute
at the California Institute of Technology. NN-EXPLORE is
managed by the Jet Propulsion Laboratory, California Institute
of Technology under contract with the National Aeronautics
and Space Administration. The authors
thank the NEID Queue
Observers and WIYN Observing Associates, as well as the California Planet Search
team supporting precise RV acquisition with the Keck~I telescope, for their dedicated
and skillful execution of the observations.

Some of the data presented in this article were obtained from the Mikulski Archive for Space Telescopes (MAST) at the Space Telescope Science Institute and can 
be accessed via doi:10.17909/19s5-fe67.

\software{\cite{IvshinaWinn2022_software}}

\bibliography{refs}{}
\bibliographystyle{aasjournal}

\appendix

\section{New transit times and RVs for WASP-4 and WASP-12}

\begin{deluxetable}{ccc}[hb!]
\label{tbl:wasp4-rv}
\tablecaption{WASP-4 HIRES relative radial velocities}
\tablehead{
  Time & Radial velocity & Uncertainty \\[-0.04in]
  (BJD) & (m/s) & (m/s)
}
\startdata
2455378.10273 & -191.35 & 2.84\\
2456152.12045 & 218.23 & 3.18\\
2456168.07375 & 199.24 & 3.14\\
2456207.94924 & -60.88 & 2.83\\
2456532.09391 & 193.72 & 2.69\\
2458662.10006 & -326.8 & 2.67\\
2458715.08168 & -31.28 & 2.74\\
2458723.06334 & 0.35 & 3.17\\
2459189.71567 & -1.21 & 2.77
\enddata
\end{deluxetable}

\begin{deluxetable}{cc}
\label{tbl:wasp4-ttv}
\tablecaption{New WASP-4 transit times}
\tablehead{
  Time of conjunction & Uncertainty \\[-0.04in]
  (BJD$_{\rm TDB}$) & (days)
}
\startdata
2460909.86707 & 0.00034 \\
2460911.20587 & 0.00034 \\
2460912.54389 & 0.00032 \\
2460913.88185 & 0.00033 \\
2460915.22090 & 0.00033 \\
2460916.55882 & 0.00032 \\
2460917.89716 & 0.00036 \\
2460919.23551 & 0.00035 \\
2460923.24982 & 0.00034 \\
2460924.58793 & 0.00035 \\
2460925.92689 & 0.00036 \\
2460928.60266 & 0.00032 \\
2460929.94090 & 0.00035 \\
2460931.27952 & 0.00034 \\
2460932.61711 & 0.00037
\enddata
\tablecomments{Based on TESS Sector 96.}
\end{deluxetable}

\begin{deluxetable}{ccc}
\label{tbl:wasp12-rv}
\tablecaption{WASP-12 NEID relative radial velocities}
\tablehead{
  Time & Radial velocity & Uncertainty \\[-0.04in]
  (BJD) & (m/s) & (m/s)
}
\startdata
2460356.79680 & 19038.4 & 7.0 \\
2460364.79782 & 18806.5 & 7.2 \\
2460370.79990 & 19217.2 & 6.9 \\
2460374.81484 & 19033.9 & 9.0 \\
2460381.69721 & 19229.3 & 7.4 \\
2460382.74067 & 19246.1 & 10.3 \\
2460383.71488 & 19234.2 & 8.0 \\
2460388.73542 & 18793.7 & 10.1 \\
2460390.77048 & 18883.9 & 10.7 \\
2460392.71535 & 19113.1 & 12.0 \\
2460415.63353 & 19142.8 & 9.8 \\
2460439.63265 & 19154.9 & 11.0 \\
2460606.94516 & 18812.6 & 6.2 \\
2460607.90961 & 18934.7 & 7.3 \\
2460608.80832 & 19141.8 & 9.4 \\
2460609.95971 & 19081.4 & 6.2 \\
2460691.69033 & 19231.5 & 8.5 \\
2460693.79940 & 19252.7 & 5.0 \\
2460694.83252 & 19260.0 & 5.9 \\
2460706.67271 & 19119.0 & 7.2 \\
2460730.63381 & 19050.3 & 5.5 \\
2460731.60757 & 18923.6 & 5.0 \\
2460732.61346 & 18834.8 & 6.1 \\
2460733.83508 & 18951.2 & 8.4 \\
2460734.67715 & 18793.2 & 9.1 \\
2460737.67836 & 19042.7 & 6.3 \\
2460739.67883 & 19239.4 & 5.0 \\
2460743.65539 & 18973.1 & 5.2
\enddata
\end{deluxetable}

\end{document}